\documentstyle[prl,aps,graphicx,twocolumn]{revtex}

\newcommand{\degree}{\mbox{\(\mathsurround=0pt{}^\circ\)}}

\begin{document}
\draft
\wideabs{
\title{Doppler cooling and trapping on forbidden transitions}
\author{T. Binnewies, G. Wilpers, U. Sterr, F. Riehle, J. Helmcke}
\address{Physikalisch-Technische Bundesanstalt, 38116 Braunschweig, Germany}
\author{T. E. Mehlst\"aubler, E. M. Rasel, W. Ertmer}
\address{Institut f\"ur Quantenoptik, Universit\"at Hannover, 30167 Hannover,
Germany}
\date{\today}
\maketitle
\begin{abstract}

Ultracold atoms at temperatures close to the recoil limit have been achieved by
extending Doppler cooling to forbidden transitions.
A cloud of $^{40}$Ca atoms has been cooled and trapped to a temperature as low
as 6 $\mu$K
by operating a magneto-optical trap on the spin-forbidden intercombination
transition.
Quenching the long-lived excited state with an additional laser enhanced the
scattering rate by a factor of 15, while a high selectivity in velocity was
preserved.
With this method more than 10 \% of pre-cooled atoms from a standard
magneto-optical trap have been
transferred to the ultracold trap. Monte-Carlo simulations of the cooling
process are in good agreement with the experiments. 
\end{abstract}
}

The preparation of ultracold atoms with temperatures far below the millikelvin
regime is essential for the progress in
many fields such as metrology, atom optics and the study of collective effects
like Bose-Einstein-Condensation (BEC).
Up to now the available techniques are restricted to a limited number of
species, mainly those that can be
cooled below the Doppler limit \cite{chu85} e.~g. by efficient polarization
gradient cooling
\cite{let88}, by velocity selective coherent population trapping (VSCPT)
\cite{asp88} or by Raman cooling \cite{kas92a}. In general, these techniques are 
only applicable to atoms with magnetic or hyperfine sub-structure in the ground
state.
Atoms with a single ground state, like the alkaline earths, often exhibit
forbidden transitions
with unique properties e.~g. for optical frequency standards
\cite{rus98,oat99,rie99}, atom
interferometry \cite{ber97c}, the study of cold collisions \cite{din99,zin00}
and possibly for achieving
BEC by all optical means \cite{cas98,mac01}.
Doppler cooling on such forbidden transitions can lead to temperatures much
lower than
the recoil limit \cite{wal89}. In strontium Doppler cooling has recently led
to phase space densities close to quantum degeneracy \cite{ido00}.
Up to now, this method failed to work for neutral atoms with ultra-narrow
(forbidden)
lines since in this case the light forces might be comparable or even smaller
than the gravitational force which makes magneto-optical trapping of atoms
difficult or impossible.

In this letter we present a novel method for effective Doppler cooling and
trapping of pre-cooled
atoms on extremely narrow transitions in three dimensions. The low
scattering rate associated with the narrow line width is artificially enhanced
by a
repumping or 'quenching' laser which de-excites the atoms via an intermediate
level.
In this case the atoms can be considered as effective two-level systems
with tunable linewidths that can be tailored during the cooling process for
optimum efficiency of the procedure.
Similar schemes were performed to more rapidly cool trapped ions \cite{die89}
and for the efficient deflection of a beam of metastable helium \cite{roo95}.
Whereas in the case of metastable helium cycling on the second transition was
simply used to increase the radiation pressure, our scheme preserves the
velocity selectivity of the narrow transition.
Our calculations for magnesium and calcium showed that this 'quench cooling' can
efficiently trap atoms released from a broad-line magneto-optical trap (MOT) and
cool them to temperatures close to the recoil limit.
We utilize this scheme for the first time to capture and cool free atoms in
three dimensions.

The temperature that can be achieved in Doppler cooling on broad lines is
determined by a balance between damping forces and the heating due to
spontaneous recoil momenta.
Here, broad lines are characterized by the linewidth of the transition $\Gamma$
being large compared to the change in the Doppler shift due to one photon
recoil, i.e. ${\frac{\hbar k^2}{m}} \ll \Gamma $ where $m$ denotes the atomic
mass and $k$ the wavevector of the exciting light.
On a two-level system this finally leads to a Doppler temperature $T_D = \frac
{\hbar \Gamma}{2 k_B}$
with the  Boltzmann's constant $k_B$.

\begin{figure}
\centerline{\includegraphics[width=6cm]{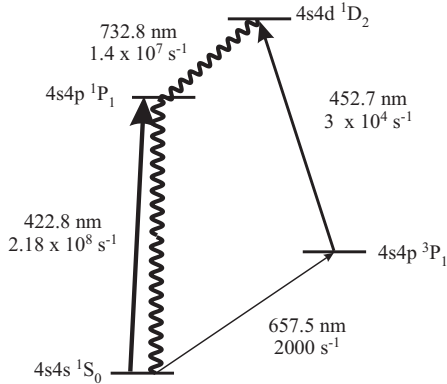}}
 \caption{Excerpt from the Ca level scheme with wavelengths and transition rates
of the relevant transitions. The $^1S_0$ - $^1P_1$ transition is used for
pre-cooling atoms in a standard broad-line MOT. The intercombination transition
$^1S_0$-$^3P_1$, quenched by the
transition $^3P_1$-$^1D_2$ forms the second stage MOT.}
 \label{scheme}
 \end{figure}

For monochromatic excitation on narrow lines, the Doppler cooling limit was
calculated to be approximately half the recoil temperature given by $k_B
T_{\mathrm{rec}} = { \frac {\hbar^2 k^2}{m}}$ \cite{cas89}.
A multichromatic excitation was proposed \cite{wal89} which would allow to reach
a temperature even below that temperature. However, narrow-line cooling is
hampered by the small force $F_{\mathrm{max}}=\frac {\hbar k \Gamma}{2}$. For
cooling of calcium on the intercombination line $^1S_0$ - $^3P_1$ (see fig.
\ref{scheme}, $\Gamma_1 = 2000~\mathrm{ s^{-1}}$) $F_{\mathrm{max}}$ is only 1.5
times the gravitational force.
To increase the force, we have reduced the lifetime of the upper level through
excitation to the $4s4d ~ ^1D_2$ state ($\lambda $ = 453~nm), from where the
atoms decay rapidly via the $4s4p$ $^1P_1$ state to the ground state.
In a different picture, the $^3P_1$ level acquires an admixture of the $^1D_2$
level (natural decay rate $\Gamma_2 = 1.4 \times 10^7 ~ \mathrm{s^{-1}}$)
through the coupling by the quench laser. Hence, for low saturation the width of
the triplet state increases proportional to the quench laser intensity.
Another possibility for quenching was demonstrated by the NIST group where
coupling to the $4s5s$ $^1S_0$ state by a 553~nm laser was used to achieve
cooling of Ca in one dimension \cite{cur01}.

In a preliminary experiment in 1-D with sequential excitation to the $^3P_1$
state and subsequent quenching we have found a quenching rate of $r_{12} \approx
1.4 \times 10^4~\mathrm{s^{-1}}$.
This is more than one order less than is calculated from the atomic data
\cite{kur88} of the quenching transition. We assume the difference to be in part
due to the uncertainty in the theoretical A-coefficient.
From our preliminary measurement we estimate a quenching rate $r_{12} \approx 3
\times 10^4~\mathrm{s^{-1}}$ for our 3--D MOT configuration, corresponding to an
average broadening factor of 15 limited by the laser power available.
In the MOT configuration, excitation to the $^3P_1$ level and quenching is
performed simultaneously. Solving the optical Bloch equations we arrive at an
effective two level system with an effective linewidth of the metastable $^3P_1$
level of $\Gamma_{\mathrm{eff}} = \Gamma_1 + r_{12}$.
Since the increased linewidth of $\Gamma_{\mathrm{eff}} \approx 2 \pi \times
5~$kHz is still small compared to the recoil shift, atoms can be lost when
kicked out of resonance by the photon recoils of one quenching cycle.
To overcome this we broaden the excitation spectrum by a harmonic frequency
modulation (peak-to-peak amplitude $\delta f_{pp} = 1.4~$MHz, modulation
frequency $f_{\mathrm{mod}}=15~$kHz) of the 657-nm laser (linewidth
approximately 500 Hz) using an acousto-optic modulator. The high-frequency end
of this frequency comb is detuned below resonance by $\Delta \approx 2 \pi
\times  280\ \mathrm{kHz}$.

With the laser power available on the intercombination transition we saturate
the transition even with the quenching laser on. With a population in the
$^3P_1$ state close to 0.4 the total scattering rate amounts to $S  \approx 1.2
\times 10^4~\mathrm{s^{-1}}$.

In the absence of a magnetic field, two cooling regimes can be distinguished:
In the Doppler regime which is effective at larger velocities, the Doppler shift
brings the atom into resonance with a counterpropagating laser beam. This leads
to a constant damping force towards zero velocity and a typical cooling time of
$t_D \approx {m v_0} / {\hbar k_1 S} = 5.6~$ms using an initial velocity $v_0$
= 1 m/s and $k_1$ being the wave vector of the 657~nm laser. Because this time
has to be small enough to prevent atoms from escaping the laser beams before
being stopped, we have increased the force by using the same circular
polarization for the cooling 657~nm beams and the quenching 453~nm laser.
Absorption of the quenching photon from the same direction as the initial photon
is favored by 6 to 1 as a result of the ratio of the Clebsch-Gordan coefficients
involved.

The second regime occurs at lower temperatures when slow atoms are not excited
by any laser and accumulate near zero-velocity. Compared to VSCPT there is still
a small off-resonant excitation probability. Therefore this regime can be
referred to as 'velocity-selective gray-state trapping'. The limit of this
cooling method depends on the detuning $\Delta$ and can be as low as a fraction
of the recoil velocity. The relevant timescale is much longer than in the
Doppler regime mentioned above as the cooling relies on random jumps to the gray
state.

The situation changes with the magnetic quadrupole field present in the MOT
configuration. Here, atoms that are in the gray state at the MOT center will
eventually move to a point where they come into resonance with one laser by the
Zeeman shift.  There they will be pumped out of the gray state and accelerated
towards the trap center until the atoms are just Doppler shifted off resonance.
Therefore the average atomic velocity will be of the order of a few recoils
\cite{kat99} independent of the detuning $\Delta$. The detuning however
determines the size of the cold atomic cloud. 

The recoil that determines the achievable velocity width is due to the
fluorescence photons at 733 nm and 423 nm from the decay of the $^1D_2$ level
(fig. \ref{scheme}) and the quench photon at 453 nm that are involved in the
final cycle before the atom gets off resonance.  The effective recoil velocity
is given by the quadratic sum of the three photon momenta and amounts to
$v_{\mathrm{rec}} \approx 3.5\ \mathrm{cm/s}$, if we neglect polarization
effects, i.e. assuming also a random direction of the absorbed quench photon.

We realized this novel cooling scheme in an experiment, where we started with a
broad-line MOT on the cooling transition $^1$S$_0$ - $^1$P$_1$ made of three
retroreflected laser beams with $\lambda=423$~nm. With a total power of 30 mW
approximately $10^7$ atoms were cooled and trapped directly from a thermal beam.
The narrow-line MOT at 657 nm was realized by three circular polarized beams
(power approx. 7~mW per beam with a diameter of 5 mm), that were parallel to the
423~nm MOT beams.
Dichroitic retroreflectors were used to combine the beams.
The quench laser beams at 453~nm were coupled into the vacuum chamber under an
angle of 22.5$\degree$ to the other cooling beams in the horizontal plane and
nearly parallel in the vertical direction.
To make efficient use of the available power ($P\approx 20~$mW) a single beam of
diameter $\approx 3$ mm was fed through the setup three times and finally
retroreflected, similar to a MOT configuration. Due to the limitations of our
vacuum chamber which was designed for an optical Ca frequency standard it was
not possible to completely match the sizes and directions of the quench and MOT
laser beams.
We expect this to limit the possible transfer efficiency in our setup.

After pre-cooling the atoms in the broad-line MOT to $v_{\mathrm{rms}} \approx
0.7~$m/s, the 423~nm laser was switched off and the magnetic field gradient was
lowered within $100~\mathrm{\mu s}$ from $60 \times 10^{-4}$ T/cm to $0.3 \times
10^{-4}$ T/cm and the 657~nm cooling laser and the quench laser were switched
on. The small gradient accounts for the much smaller acceleration that can be
reached by cooling on the narrow line, even with quenching.
After the cooling time the 657~nm laser was switched off while the quench laser
remained on to repump most of the atoms to the ground state.  To reduce the
background from fluorescence of decaying atoms that have not been repumped, we
waited for 0.5~ms to 1~ms before measuring the velocity distribution.

\begin{figure}
\centerline{\includegraphics[width=8cm]{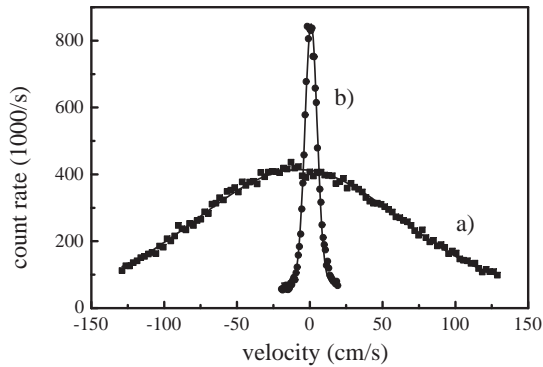}}
 \caption{Velocity distribution measured by the excitation spectrum of the
Doppler broadened intercombination transition. The distribution after cooling in
a usual broad-line MOT (a) shows an rms velocity of 71 cm/s ($T = 2.4$ mK )
which is reduced to 4.3 cm/s ($T = 9~\mu$K) after 25 ms of second stage cooling
(b). Comparing the areas of the two distributions gives a transfer efficiency of
12 \%.}
 \label{distribution}
 \end{figure}

The velocity distribution of the atomic ensemble was obtained by measuring the
Doppler broadening of the intercombination transition. Using long excitation
pulses the corresponding interaction-time broadening could be neglected. To
avoid Zeeman broadening of the transition the magnetic field was changed from a
quadrupole to a homogeneous field.
Fig. \ref{distribution} shows a typical velocity distribution detected from the
fluorescent decay of the $^3P_1$ state before and after the cooling on the
quenched intercombination line.
To determine the efficiency of the cooling process we compare the areas of the
two curves giving a transfer efficiency of $\eta_N =$ 12 \% in the case of fig.
\ref{distribution}.
Due to the integration over all transverse velocities the increase in count rate
by a factor of 2 shown in fig. \ref{distribution}, does not reflect the increase
in the number of atoms $\eta_v$ with $v \approx 0$. This increase is calculated
from
$\eta_v = \eta_N ({v_{\mathrm{rms,~423}}}/{v_{\mathrm{rms,~657}}})^3$
as $\eta_v \approx$ 500.

\begin{figure}
\centerline{\includegraphics[width=8cm]{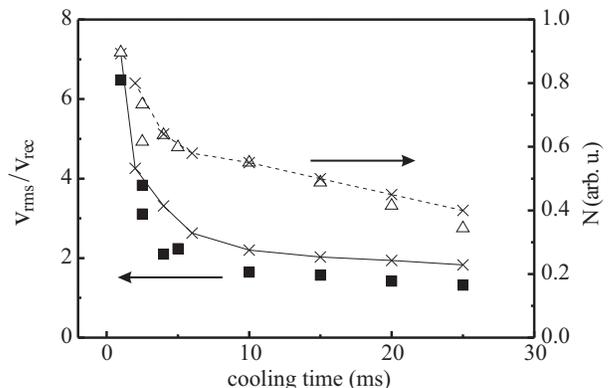}}
\caption{Atom numbers (triangles) and root-mean-square velocities
$v_{\mathrm{rms}}$ (squares) in units of the effective recoil velocity
$v_{\mathrm{rec}}= 3.5$ cm/s after second stage cooling versus the cooling time.
The crosses connected by lines show the results of Monte-Carlo
simulations using the experimental parameters. The simulated atom number is
scaled to fit the measurement.}
\label{time}
\end{figure}

The temperature as a function of the cooling time (fig. \ref{time}) shows an
exponential decay with a time constant of 1.4~ms until the rms velocity is
reduced to a few times the effective recoil velocity $v_{\mathrm{rec}}$. For
longer cooling times a further slow reduction of the velocity width is observed.
A temperature of $(5.6 \pm 0.4)~\mu$K after 100~ms of cooling has been observed,
which corresponds to an rms velocity of 3.4 cm/s that is close to the calculated
effective $v_{\mathrm{rec}}$. 

The time dependence of the total number of atoms as deduced from the area of the
measured Doppler curves (fig. \ref{time}) shows first a fast loss due to atoms
that could not be captured by the narrow-line MOT and then an almost linear
decrease in atom number with cooling time.
From absorption images of the ultracold atomic cloud after a cooling time of 15
to 20 ms we obtain rms radii of 0.8 mm and 0.4 mm in the horizontal and vertical
directions, respectively.

\begin{figure}
\centerline{\includegraphics[width=8cm]{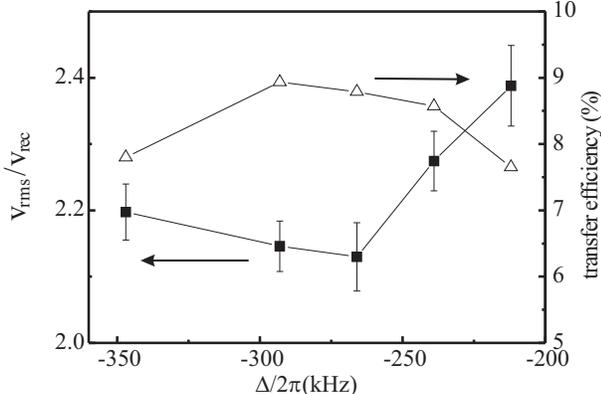}}
\caption{Root-mean-square velocity and transfer efficiency after 25 ms of second
stage cooling versus the detuning of the 657~nm MOT laser. The bars indicate the
statistical uncertainty of $v_{\mathrm{rms}}$ from the fits to the measured
velocity distributions.}
\label{detuning}
\end{figure}

In our present setup significant variations of the final temperature and the
transfer efficiency occur over the day as can be seen from the different scales
of fig. \ref{time} and fig. \ref{detuning}. We attribute this to slight changes
in the alignment and the detuning of the quench laser.

We have modelled the cooling and trapping by a semiclassical 3-D Monte-Carlo
simulation. The temporal dependence of the temperature is in good agreement with
the experimental results also showing a fast reduction and a further slow
decrease of the rms velocity approaching the recoil velocity for long times. The
calculated transfer efficiency typically amounts to 60 \%, far above the
experimentally observed 12 \%, probably due limitations of our setup as
mentioned above.
The slow loss of atoms also shows up in the Monte-Carlo simulations, where it
was identified as caused by atoms moving out of the intersection region of the
laser beams. Similar temperatures near the recoil limit have been obtained with
simulations for the narrow-line cooling of magnesium on the intercombination
transition $3s^2$ $^1S_0$ - $3s3p$ $^3P_1$ and quenching using the $3s3p$
$^3P_1$ - $3s4s$ $^1S_0$ transition.

The detuning dependence of the final temperature is shown in fig.
\ref{detuning}. The velocity reaches a minimum around 270 kHz and is nearly
constant for larger detunings, as discussed above. The increase at lower
detunings is probably due to excitation in the wings of the excitation profile.
The transfer efficiency shows a maximum at approximately the same detuning, as
for smaller detunings the capturing probability is reduced. For larger detunings
the loss increases because the trap becomes larger than the intersection region
of the laser beams.

This quench cooling scheme has the distinct advantage that the scattering rate
can be tailored for different purposes by adjusting the power of the quench
laser, e.g. using a high scattering rate for efficient capturing and then
lowering the rate to reach the lowest temperatures.
For use in an atomic frequency standard ultracold atoms at moderate density are
desired. This can be
reached by completely switching off the magnetic field and finally cooling the
atoms in an optical molasses to reach even sub-recoil temperatures.  On the
other hand, to enable efficient filling of an optical dipole trap one would
increase the magnetic field gradient after a few milliseconds to reach a high
phase-space density.

The presented results open the way to a wide variety of experiments including
the formation of quantum
degenerate ensembles by all-optical means. This might prove particularly useful
in the case of
Ca where calculations \cite{mac01} indicate a positive scattering length.

In conclusion, we have presented a cooling method applicable to atoms with
narrow or forbidden
transitions. We have demonstrated this method by cooling a $^{40}$Ca atomic
ensemble in three dimensions to a residual rms-velocity close to one effective
recoil. Our scheme is applicable to arbitrarily narrow atomic transitions that
can be artificially broadened and, hence, does not suffer from the associated
small forces.

We thank J. Keupp and S. Baluchev for technical assistance.
The work was supported by the Deutsche Forschungs\-gemeinschaft under SFB 407
and by the European Commission through the Human Potential Programme (Cold Atoms
and Ultra-precise Atomic Clocks; CAUAC).

\end{document}